\documentclass{sig-alternate}

\usepackage[utf8]{inputenc}
\usepackage{CJKutf8} 
\usepackage{booktabs}
\usepackage[hyphens]{url}
\usepackage{ucs}
\usepackage[encapsulated]{CJK}

\def\alignauthor[#1]{
\end{tabular}%
  \begin{tabular}[t]{p{#1}}\centering}%

\begin{document}
\conferenceinfo{WebSci}{'15 Oxford, UK}

\toappear{\copyright~Liao, H.T., Fu, K.W., Hale, S.A. 2015. This is the authors' version of the work. It is posted here for your personal use. Not for redistribution. The definitive version was published in WebSci '15, http://dx.doi.org/10.1145/2786451.2786486.}

\title{How much is said in a microblog?\\A multilingual inquiry based on Weibo and Twitter}

\numberofauthors{3} 
\author{
\alignauthor[0.31\textwidth]
Han-Teng Liao\\
       \affaddr{Oxford Internet Institute}\\
       \affaddr{University of Oxford}\\
       \email{hanteng@gmail.com}
\alignauthor[0.34\textwidth]
King-wa Fu\\
       \affaddr{Journalism and Media Studies Centre}\\
       \affaddr{University of Hong Kong}\\
       \email{kwfu@hku.hk}
\alignauthor[0.31\textwidth]
Scott A.~Hale\\
       \affaddr{Oxford Internet Institute}\\
       \affaddr{University of Oxford}\\
       \email{scott.hale@oii.ox.ac.uk}
}

\maketitle
\begin{abstract}
This paper presents a multilingual study on, per single post of microblog text, (a) how much can be said, (b) how much is written in terms of characters and bytes, and (c) how much is said in terms of information content in posts by different organizations in different languages. Focusing on three different languages (English, Chinese, and Japanese), this research analyses Weibo and Twitter accounts of major embassies and news agencies. We first establish our criterion for quantifying ``how much can be said'' in a digital text based on the openly available Universal Declaration of Human Rights and the translated subtitles from TED talks. These parallel corpora allow us to determine the number of characters and bits needed to represent the same content in different languages and character encodings. We then derive the amount of information that is actually contained in microblog posts authored by selected accounts on Weibo and Twitter. Our results confirm that languages with larger character sets such as Chinese and Japanese contain more information per character than English, but the actual information content contained within a microblog text varies depending on both the type of organization and the language of the post. We conclude with a discussion on the design implications of microblog text limits for different languages.
\end{abstract}




\section{Arbitrary character limits?}
Microblogging platforms are distinguished from traditional blogging platforms by having a limit on the length of posts. 
Such length limitations 
are reported to lower the time and thought required to create new content and allow for faster and more timely exchange of information \cite{ebner2008,java2007}. 
Most studies considering the effects of length limitations on the user experience of microblogging have examined English-language content, but there is ample reason to believe that universal length limitation have different effects on people writing content in different languages. It has been reported by major media such as BBC \cite{hewitt_has_2012} and the Atlantic \cite{rosen_how_2011} that one can express much more content in languages other than English within a  given character limit, most notably the 140-character limit of Twitter and the 140-byte limit of Short Message Service (SMS, or text messages). Chinese \cite{chan_microblogs_2010} and Japanese \cite{summers_whats_2010} are often cited as examples of ``more expressive'' languages within such space limits.

However, a definitive answer using systematic and quantifiable methods is yet to be provided to the question of how much more expressive a given language is within such length limitations. Neubig and Duh \citeyear{neubig_how_2013} provide the only academic work on this subject and use an information-theoretic approach. They find that Chinese and Japanese are the most expressive languages per character, but do not use parallel corpora (i.e., the same information in multiple languages) in their work. Apart from academic scholarship, the issue of language expressiveness has received much attention in the popular press with the BBC reporting that 140 Chinese characters amounts to 70 to 80 English words \cite{hewitt_has_2012}. Bloggers have also weighed in on the debate with one blogger writing that 140 Chinese characters could contain five times more content than the same number of English characters \cite{ruby_twitter_2012} and another blogger claiming that ``140 Chinese characters is more like 500 characters on Twitter.com'' \cite{dugan_140_2011}. By machine translating foreign-language content from Twitter from ``a few users,'' IT Consultant Ben Summers reported on his blog that Japanese tweets could contain information that would take up to 260 English characters to express \cite{summers_whats_2010}. These differing estimates and the lack of academic scholarship on this topic motivate our paper.

The imposition of length constraints has profound effects on how platforms are used and hence the user experience of these platforms.
The effects of length constraints on users are not new or unique to microblogging platforms. The user experience of SMS has received much attention, and scholars have found that the limitations resulted in specific language practices including specific abbreviations (e.g., b4, 2day) as well as more contractions in comparison to instant messaging \cite{ling_text_2007}. It is difficult to disentangle length and input device limitations when comparing instant messaging and SMS \cite{ling_text_2007}, but one study found SMS messages sent via Internet-connected PCs were longer than those sent directly from mobile (feature) phones \cite{grinter2001}. Finally, Grinter et al.~\cite{grinter2001} found that the length limitations of SMS allowed teenagers to forego conversational conventions and reduce the overall time spent on interactions.

In this paper, we attempt to answer the general questions of how space/length restrictions affect (a) how much can be said, (b) how much space is used, and therefore (c) how much is actually said within imposed space limitations and how these factors differ across languages. These basic questions are central to any cross-lingual assessment of the impact of length limitations on microblog posts and hence the user experience of people on such services and have important ramifications for designers of microblogging platforms as well as people using such platforms.

\subsection{Reasons for length limitations}
Early character limits were often imposed as a result of technical limitations. For example, the character limit of text messages (SMS) was the result of an engineering choice to send SMS messages via the existing signaling channels of the GSM system (but with a lower priority than other signaling messages) \cite{hillebrand2002}. After including header information, the existing channel left 140 octets (bytes) available for the text of an SMS message. Using 7-bit character encoding, 160 Latin characters were made to fit in this space \cite{ETSI-TS-123-040}. Hillebrand, an early pioneer of SMS, reasoned that 160 characters would be sufficient through an examination of the number of characters (including spaces and punctuation marks) for random short paragraphs \cite{gayomali_text_2012,sun_cross-cultural_2012}. Although we cannot be certain which language he used for this exercise, it is likely to have been English or German. If another language (particularly a language with a non-Roman alphabet such as Japanese or Chinese) had been used, Hillebrand may have reached a different conclusion about the feasibility of using the existing signaling channels of the GSM system and ultimately developed SMS in a different manner with a different character limit.

The continuation of character limits on modern microblogging platforms is less a case of technical limitations and more often a design choice aiming to cultivate a certain user experience. Best practices published by Twitter in 2013 mentioned the 140-character limit and stated that ``creativity loves constraints and simplicity is at our core.'' \cite{twBestPractices}. Character limitations may have very different effects in different languages as cross-platform and cross-lingual analysis of microblogs suggests language is an important factor for researchers to consider [e.g., \cite*{fu_assessing_2013,hale2012,hale2014,hong2011,liao2014}].

When Twitter launched, it set a limit of 140 characters for posts so that a Twitter post would fit within a single SMS message and leave 20 characters for usernames and other commands \cite{gayomali_text_2012,milian_why_2009}. It should be noted, however, that only for 140 English characters would this have ever been the case as SMS itself has a hard limit of 140-bytes within a single message as previously stated. Thus, a single SMS message can only accommodate 70 Chinese or Japanese characters. The limits imposed on Twitter, however, are character limits and not byte limits \cite{twCharacterCounting}, and thus, it is possible to send a tweet with 140 Chinese characters even though such a tweet would never have fit within one SMS message. Now that most Twitter traffic comes from native apps on smartphones, which have no technical limitation on the length of messages, the continued imposition of the 140-character limit is due less to technical reasons and more to user-experience design choices.

When Sina Weibo launched in mainland China it also imposed a limit on the length of posts. This limit is often reported to also be 140 characters [e.g., \cite*{dugan_140_2011}], but the data we collected in this study revealed several posts with more than 140 characters. Our experimentation with the Sina Weibo interface indicates that Sina Weibo likely imposes a byte limit and not a character limit.
We found that posts on Sina Weibo can be up 140 Chinese characters or 280 English/Latin characters.\footnote{This suggests Sina Weibo imposes a byte limit with a character encoding such as GBK that uses one byte to store most English/Latin characters and two bytes to store other characters including most Chinese characters.} In other words, Sina Weibo imposes a varying limit on the number characters a message may contain depending on the number of bytes required to store each character. In practice, this means that messages in English on the platform could be twice as long as messages in Chinese, with most other languages having maximum lengths somewhere between these endpoints depending on the frequency of accented and other special characters in the language.

Apart from the question of the actual limits imposed on different communication platforms, marketers and other users of microblogging platforms have asked what the ``ideal'' message length is for driving strong engagement, with different numbers reported for different platforms including Twitter, Facebook and Google Plus \cite{lee_proven_2014}.  For example, it is reported and recommended by various industry research organizations that the ideal length of a tweet is either 100 or 71--100 characters \cite{lee_proven_2014, track_social_track_2012}. It is also reported that the ideal Facebook post be around 40 characters while the Google Plus posts should be around 60 characters \cite{track_social_track_2012}, despite these platforms allowing much longer posts in practice \cite{techcruch_gplus_fb}.
However, despite the marketing and industry interest in the factor of microblog post length for engagement optimization, there is little research on length across different languages and/or different platforms. In other words, we need to fill the gap in both research and design with cross-cultural considerations \cite{sun_cross-cultural_2012} regarding both the limits and actual practices regarding the length of posts.

\subsection{A parallel corpora approach}
Corpus linguists have partially addressed the question on how much can be said in different languages using parallel corpora.
Before Twitter came into existence, the Director of Linguistic Data Consortium (a major linguistic research data keeper) conducted a well-designed comparison of Chinese and English using the LDC parallel Chinese\slash{}English corpora, producing ratio results ranging from 1.96 to 2.27 for texts and 1.19 to 1.24 for compressed (gzipped) files  \cite{liberman_language_2005}.
The underlying idea behind Liberman's research design was straightforward. By comparing the actual storage space of texts of the same content in different languages, one can see how much ``space'' is required to store/convey the same content. This idea is expressed by the equation below:

\begin{equation}
S_{lang_A} \times IC_{lang_A} = S_{lang_B} \times IC_{lang_B}
\label{eq:liberman}
\end{equation}

$S_{lang}$ denotes the ``space'' units required to store the content in language $lang$. $IC_{lang}$ denotes the amount of ``information content'' per space unit in language $lang$. The equation holds only for parallel corpora, which by definition contain the same content in each language (i.e., where $lang_A$ and $lang_B$ contain the same information written in different languages).

While Liberman's research is systematic and addresses the computational questions of information storage, it is of limited application to modern social media data for two reasons.
First, the amount of storage space required to store information in different languages depends on the character encoding used. Liberman's work used the Chinese national standard GB-2312 and not the international Unicode encodings, which have become more common for multilingual websites and applications.
Second, the data Liberman used was formal and legalistic in nature. The ratios between such formal text may not apply to the more conversational and informal text commonly found on modern social media platforms.

\section{Methods and data selection}
We propose a systematic step-wise approach that can be extended to cover more platforms, languages, and types of microblogs, with the aim to answer our three main research questions. We choose to focus on news and diplomacy organization accounts because we expect the length of messages would be integral to the overall communication strategies, including in potential cross-lingual or cross-cultural scenarios. The data selection in this paper is limited to two platforms, three languages, and the most recent posts of 54 microblogs. However, as the first multiplatform and multilingual study its contributions are important and lay the foundation for future work with additional account types, languages, and platforms.

The following sections describe the three steps of the research process, each of which answers one of our research questions concerning (a) how much can be said, (b) how much is typed and posted, and (c) how much is said in different languages by different organizations. First, we calculate the cross-lingual ratios of information content.

\subsection{Calculating the cross-lingual ratios of information content based on UDHR and TED talks}
Following the approach of Liberman's corpus linguistic study \cite{liberman_language_2005}, we propose a generic research design to measure the ratios of ``information content'' in different languages.

The ratio of information content per space for language $B$ to language $A$ can be derived from Equation \ref{eq:liberman} to produce the ratio given in Equation \ref{eq:ratio}. This is  the inverse ratio of ``space'' required to store the same content when a parallel corpus is used. In other words, if more space is required to store the same content for language $B$ than language $A$, the ratio value will be less than one, indicating the information content per space unit of language $B$ is smaller than that of language $A$.

\begin{equation}
	ratio(lang_B,lang_A) = \frac{IC_{lang_B}}{IC_{lang_A}}=\frac{S_{lang_A}}{S_{lang_B}}
    \label{eq:ratio}
\end{equation}

The potential of using Web content as a parallel corpus for research has been proposed \cite{resnik_web_2003} and executed \cite{li_mining_2008}.
We use a parallel corpus formed from human-translated user-generated content for several reasons. First, the content is open and freely available providing for easier replication in comparison to conventional parallel linguistic corpora, which often require license fees to use. Second, the human translation of the content usually provides better quality text than corpora formed with machine translation of open content. Third, most user-generated content is generally more up-to-date and contemporary than that of conventional corpora and is closer in style to the text used on microblogging platforms and thereby provides a more suitable basis for the research on microblogging.

We analyze two corpora to understand how corpus selection influences our results. The first corpus we use is the UDHR in Unicode Project, which provides translations of the Universal Declaration of Human Rights (UDHR) to demonstrate the use of Unicode for multilingual environments. Because of the normative, universal and semilegal status of UDHR, the Unicode translation project is the ``most translated text'' \cite{un_worlds_2007}. Thus, the project can provide parallel corpora that cover the most languages in the world for comparison.

Our second parallel corpus is formed from the TED Open Translation Project. This project is led by a well-known, Internet-friendly organization and uses professional human translation service to kick-start the crowd-sourced translation of video subtitles. The translated video subtitles form a multilingual corpora that is comparatively closer in style to informal, online communication. Of course speeches on TED.com are still different from online expressions such as microblog posts, but they are closer to online expressions when compared to legal or governmental data that is commonly used in corpus linguistics because of its institutional availability.

To calculate the ratios of information content across languages, we downloaded the texts from the UDHR in Unicode Project and subtitles for all of the 1,847 TED talks available from TED.com (a complete sample of all videos available as of March 16, 2015). We have downloaded the available transcripts of four language versions: English, Japanese, simplified Chinese, and traditional Chinese. Our full analysis of the TED talks include 1,522 videos from all the videos available because 209 videos did not have transcripts available in all four languages and a further 116 videos had extremely short lengths (these were mostly performance videos).

The texts were parsed and the number of characters in each video in each language was determined. For the UDHR datasets, we use each paragraph as a unit to calculate the respective ratios across languages and then produce basic descriptive statistics. For the TED datasets, we use each speech (i.e., all the subtitles  for one video) as a unit.

\subsection{Measuring the text lengths of posts from selected accounts from Twitter and Weibo}
To measure the text lengths of microblog posts, we examine the posts by 54 news and embassy organizations. 

For Twitter, we selected 36 accounts that post either in English, Japanese, or Chinese. The most recent 200 tweets for each account were collected through the Twitter REST API on 22 January 2015. For Weibo, we collected samples from a data intermediary called Weiboscope \cite{fu_assessing_2013,fu_reality_2013} that provided uncensored and randomly sampled datasets from Weibo. The data was first collected by making SQL requests to the database for messages posted between 1 January and 22 January 2015. To ensure the number of posts per account was large enough, we only considered accounts having more than 50 posts. Two Weibo accounts that are owned by the World Bank and the Economist (with respective screen names ``\CJK{UTF8}{gbsn}{世界银行}'' and ``\CJK{UTF8}{gbsn}{经济学人集团}'') were thus excluded, leaving us with 18 accounts for analysis. The first column of Table \ref{tbl:accounts} lists the screen names of the Twitter and Weibo accounts analyzed.

\begin{table}
\begin{center}
\begin{tabular}{lll}
\textbf{Twitter}\\
\toprule
Screen name	&	Language	&	Type\\
\midrule
ft	&	English	&	news\\
ftchina	&	English	&	news\\
KyodoNewsENG	&	English	&	news\\
wsj	&	English	&	news\\
xinhuanetnews	&	English	&	news\\
47news	&	Japanese	&	news\\
asahi	&	Japanese	&	news\\
bbcjapan	&	Japanese	&	news\\
mainichiJapaneseews	&	Japanese	&	news\\
nikkei	&	Japanese	&	news\\
peopledailyjp	&	Japanese	&	news\\
sankei\_news	&	Japanese	&	news\\
WSJJapan	&	Japanese	&	news\\
asahi\_shinsen	&	Simplified Chinese	&	news\\
bbcchinese	&	Simplified Chinese	&	news\\
china\_kyodonews	&	Simplified Chinese	&	news\\
chinesewsj	&	Simplified Chinese	&	news\\
djy\_cn	&	Simplified Chinese	&	news\\
dw\_chinese	&	Simplified Chinese	&	news\\
people\_cn	&	Simplified Chinese	&	news\\
voachina	&	Simplified Chinese	&	news\\
UKinJapan*	&	English	&	embassy\\
UN	&	English	&	embassy\\
usembassytokyo*	&	English	&	embassy\\
worldbank	&	English	&	embassy\\
ChnEmbassy\_jp	&	Japanese	&	embassy\\
Embassy\_ItalyJP	&	Japanese	&	embassy\\
IcelandEmbTokyo	&	Japanese	&	embassy\\
IsraelinJapan	&	Japanese	&	embassy\\
koreanemb\_japan	&	Japanese	&	embassy\\
NLinJapan	&	Japanese	&	embassy\\
UKinJapan*	&	Japanese	&	embassy\\
UKRinJPN	&	Japanese	&	embassy\\
usembassytokyo*	&	Japanese	&	embassy\\
worldbanktokyo	&	Japanese	&	embassy\\
france\_in\_china	&	Simplified Chinese	&	embassy\\
UNRadioChinese	&	Simplified Chinese	&	embassy\\
usa\_china\_talk	&	Simplified Chinese	&	embassy\\

\\
\textbf{Weibo}\\
\toprule
Screen name	&	Language	&	Type\\
\midrule
\CJK{UTF8}{gbsn}{人民网}	&	Simplified Chinese	&	news\\
\CJK{UTF8}{gbsn}{日本共同社}	&	Simplified Chinese	&	news\\
\CJK{UTF8}{gbsn}{韩国中央日报}	&	Simplified Chinese	&	news\\
\CJK{UTF8}{gbsn}{半岛电视台AlJazeera}	&	Simplified Chinese	&	news\\
\CJK{UTF8}{bsmi}{ETtoday新聞雲}	&	Simplified Chinese	&	news\\
\CJK{UTF8}{gbsn}{FT中文网}	&	Simplified Chinese	&	news\\
\CJK{UTF8}{gbsn}{华尔街日报中文网}	&	Simplified Chinese	&	news\\
\CJK{UTF8}{gbsn}{福布斯中文网}	&	Simplified Chinese	&	news\\
\CJK{UTF8}{gbsn}{路透中文网Reuters}	&	Simplified Chinese	&	news\\
\CJK{UTF8}{gbsn}{联合国}	&	Simplified Chinese	&	embassy\\
\CJK{UTF8}{gbsn}{美国驻华大使馆}	&	Simplified Chinese	&	embassy\\
\CJK{UTF8}{gbsn}{加拿大大使馆官方微博}	&	Simplified Chinese	&	embassy\\
\CJK{UTF8}{gbsn}{韩国驻华大使馆}	&	Simplified Chinese	&	embassy\\
\CJK{UTF8}{gbsn}{以色列驻华使馆}	&	Simplified Chinese	&	embassy\\
\CJK{UTF8}{gbsn}{俄罗斯驻华大使馆}	&	Simplified Chinese	&	embassy\\
\CJK{UTF8}{gbsn}{韩国驻华大使馆}	&	Simplified Chinese	&	embassy\\
\CJK{UTF8}{gbsn}{以色列驻华使馆}	&	Simplified Chinese	&	embassy\\
\CJK{UTF8}{gbsn}{俄罗斯驻华大使馆} &	Simplified Chinese	&	embassy\\
\bottomrule
\end{tabular}
\caption{Twitter and Weibo accounts analyzed in this study}
\label{tbl:accounts}
\end{center}
\end{table}

With the help of human readers and our own language identification algorithms, we found that almost all accounts used one language. We coded each account accordingly with their respective language codes. We found that two Twitter accounts posted content in both English and Japanese (``UKinJapan'' and ``usembassytokyo''). We collected 100 posts in each language for each of these accounts, and analyze the content in each language separately. Each of these accounts is marked with an asterisk (*) in Table \ref{tbl:accounts}.

We coded each organization account as either \emph{embassy} or \emph{news} (the \emph{type} column of Table \ref{tbl:accounts}). A small amount of gray area exists between these two categories. For example, news organizations such as China's People's Daily (``people\_cn'') may also be media organs of state governments. Other organizations, such as the UN and World Bank, are not strictly  embassies. Nevertheless, these international organizations have their political significance in diplomacy and in some ways their use of microblogs will be similar to the use of microblogs by embassies. At the very least, these embassy and embassy-like organizations provide a contrast in type to news organizations.
The selected accounts in Table \ref{tbl:accounts} thus contain a mixture of microblogs written in different languages and belonging to different types of organizations. Although the selection of accounts and posts is not random, the selected accounts cover major news and embassy organizations across three major languages on two major microblogging platforms.

To measure the text lengths, we first remove all hyperlinks in the microblog posts and calculate the number of Unicode characters per post.%
\footnote{We remove URLs from posts before making our length comparison. Hong et al.~\cite{hong2011} found that tweets in different languages included URLs at different frequencies, but within our restricted set of embassies and news organizations, we find that URLs are included within tweets at a similar rate between the three languages we analyze. The majority of tweets we analyze in each language contained exactly one URL (67\% of the Chinese tweets, 69\% of the English tweets, and 72\% of the Japanese tweets that we analyze had exactly one URL). The analysis presented here strips URLs from all tweets and compares the results, but repeating the analysis with only tweets that contained exactly one URL yields nearly identical findings.}
Based on this measurement we derive the average length in characters (i.e., the mean value of characters per post) for each account (we ignore the differences between single-byte and multi-byte characters as Twitter does in enforcing its character limit \cite{twCharacterCounting}). With the average length values across different platforms, languages, and types of organizations, we should find out whether and how such character lengths vary, thereby answering the question of how much is typed and posted.

\subsection{Estimating the relative information content in a microblog post for cross-lingual comparison}
After we calculate the ratios of information content and the lengths, we can then derive an estimation of relative information content (RIC) as given in Equation \ref{eq:ric}.

\begin{equation}
RIC_{lang_A} = \frac{S_{lang_B}}{ratio(lang_B,lang_A)}=S'_{lang_A}
\label{eq:ric}
\end{equation}

The main purpose of the Equation \ref{eq:ric} is to allow for cross-lingual comparisons on an equal basis. Using language $A$ as the baseline, the relative information content can be derived by dividing the value of the number of space units ($S_{lang_B}$) used when the content is expressed in language $B$ with the ratio of information content ($ratio(lang_B,lang_A)$). The formula produces the equivalent space that would be required to expressed the same content in language $A$ ($S'_{lang_A}$). We present results where space is measured either as characters or as UTF-8 bytes.

\section{Findings}
The findings are presented in three subsections, each answering one of our three main research questions.
The findings of the first subsection confirm that Chinese and Japanese contain more information per character. The findings of the second subsection show a mixed picture on the length of microblog posts by different organizations in different languages. Combining the results from the first two subsections, the third and final subsection presents the derived measurement of the relative information content expressed in microblog posts by different organizations in different languages.

\subsection{How much can be said: UDHR and TED talks}

\begin{figure}
\begin{center}
	\includegraphics[width=\columnwidth]{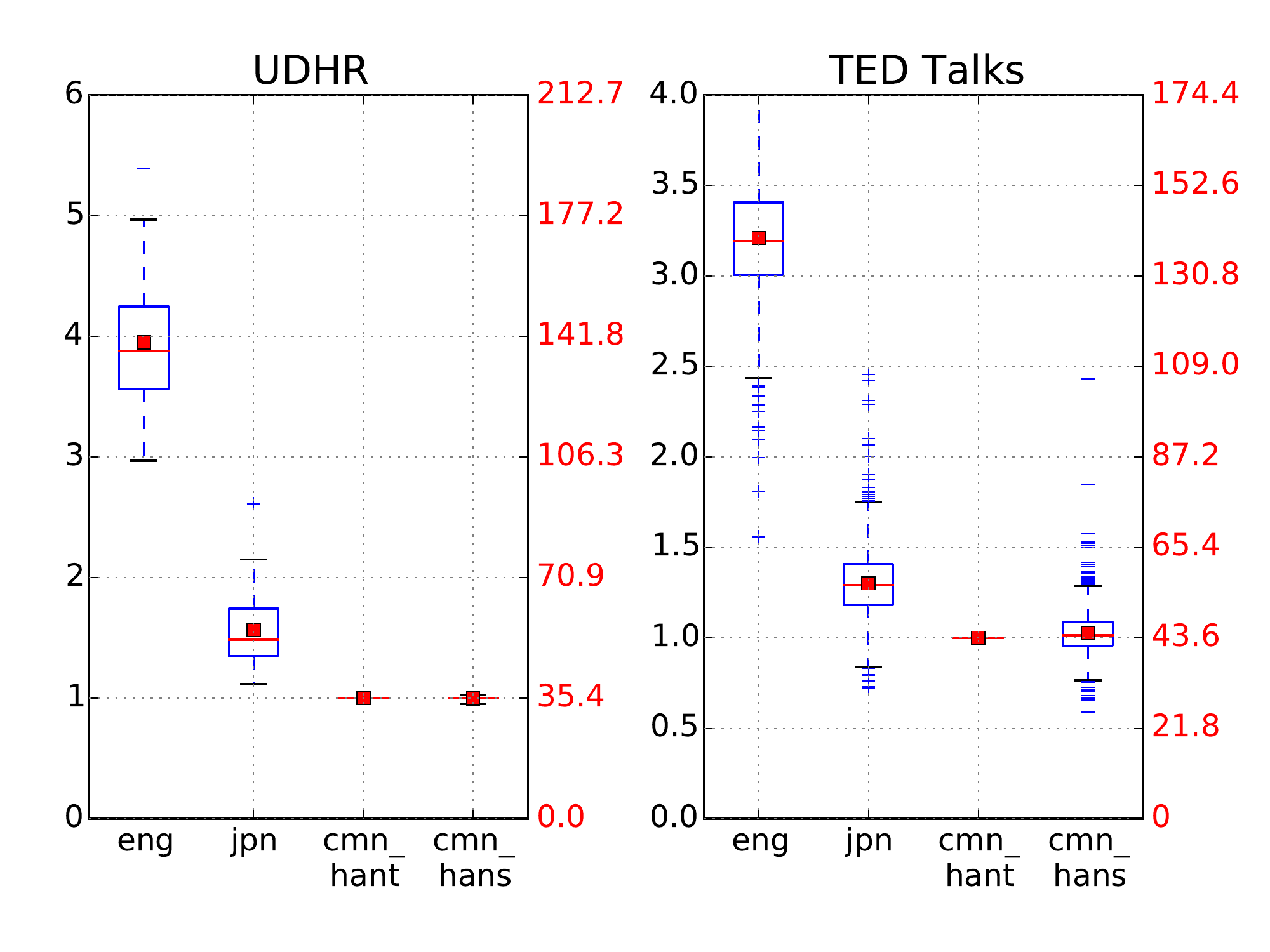}
	\caption{Relative ratio of characters in English (eng), Japanese (jpn), and Chinese using simplified characters (cmn\_hans) required to express the same content compared to Chinese using traditional characters (cmn\_hant) as the baseline.}
	\label{fig:chars}
\end{center}
\end{figure}

Figure \ref{fig:chars} shows the outcomes of the research on the UDHR texts and the TED talks respectively using box plots. Both sets of box plots show that indeed, per Unicode character, the Chinese language can express the same idea with fewer characters.
Using Chinese written with traditional characters as the baseline, the set of box plots to the left shows for the same paragraph content of UDHR that it takes on average nearly four times as many characters in English to express the same content (the mean value of 3.95 is shown by the red rectangular box in the \emph{eng} column).
Similarly, it takes only about 1.6 as many characters to express the same content in Japanese.
There is no significant difference in the number of traditional or simplified characters needed to write the same content in Chinese.
If the type of language used in the UDHR was typical of the language used in microblog posts, these findings would indicate that a tweet (or a Sina Weibo post) of 140 Chinese characters could convey nearly four times as much information as an English message of the same character length and 1.6 times as much information as a Japanese message of the same character length.

In order to understand the impact of the specific parallel corpus used on these calculations, we repeat the same analysis using the speech content of TED talks and show the results in the right set of box plots in the same figure (Figure \ref{fig:chars}).
This analysis yields slightly lower numbers, showing that it takes on average about 3.2 times the number the characters to express the same content in English in comparison to Chinese. Similarly, it takes about 1.3 times as many characters to express the same content in Japanese in comparison to Chinese. Once again, there is no significant difference between the number of simplified or traditional characters needed to express the same content in Chinese.

Relating these findings to microblogs, we can calculate the equivalent number of characters in each of our languages compared to 140 characters in English. These results are shown by the scale on the right y-axis for each set of box plots where the mean for English is set to 140 characters.
Based on the UDHR outcomes, we find that 140-character worth of English content can be expressed in 35.53 Chinese characters or 55.71 Japanese characters. Similarly 140 characters worth of English content of a TED talk can be expressed in 43.61 Chinese characters or 56.70 Japanese characters. The differences between our two corpora are likely due to the nature of the content of each corpus, as the language used in the UDHR is much more formal and legalistic in comparison to the transcripts of TED talks.%
\footnote{One limitation of using TED talks is that the vast majority of talks are given in English and other language transcripts are translations of this English source. There may be differences in how closely transcripts in each language follow the spoken dialogue, but we find that filler words (e.g., uh, er, and um) are rarely transcribed in any language, including English. Future work may consider developing a more balanced corpus by using movie subtitles with a variety of source languages.}

\begin{figure}
\begin{center}
	\includegraphics[width=\columnwidth]{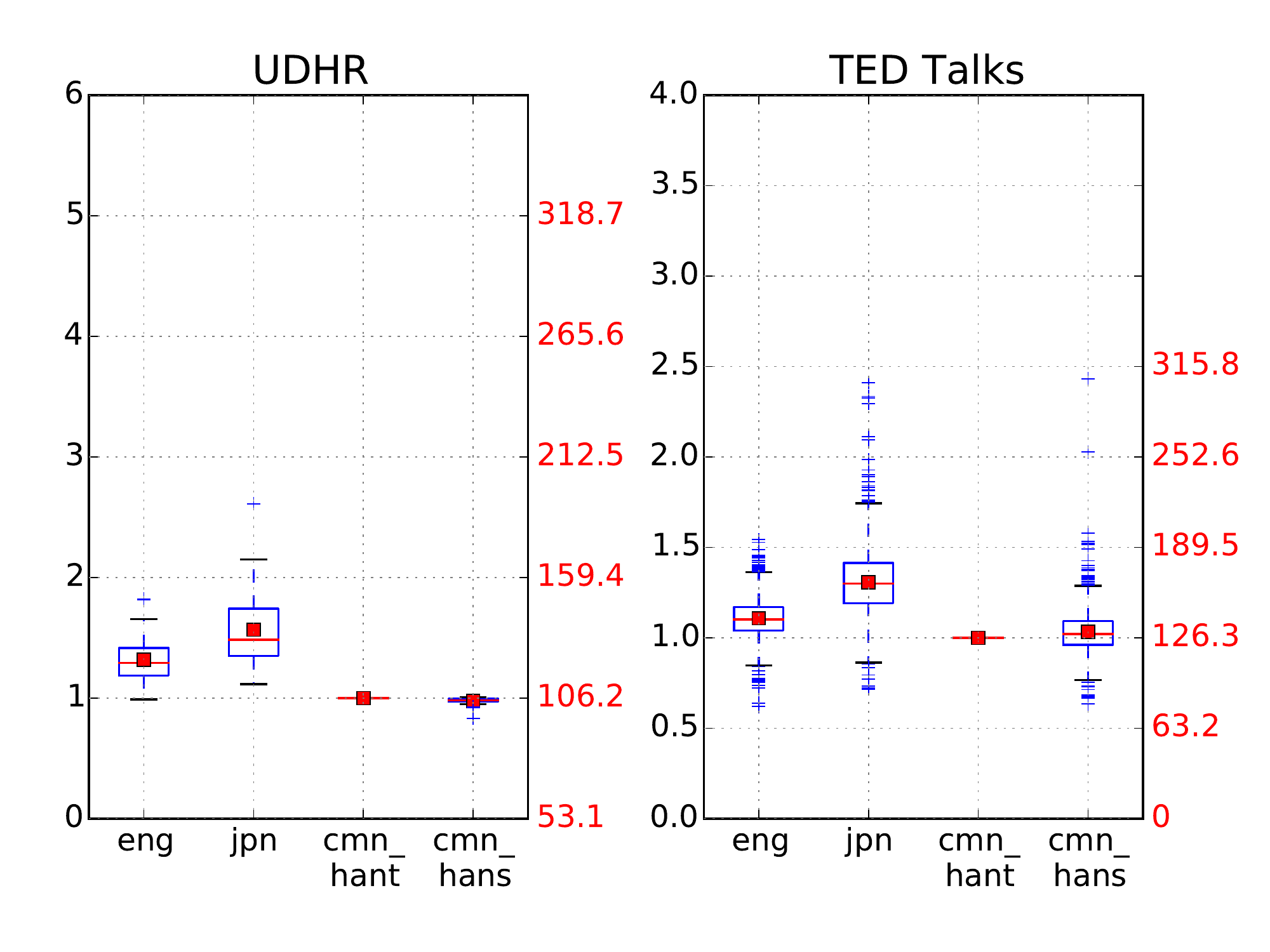}
	\caption{Relative ratio of bits (when the UTF-8 encoding scheme is used) required to express the same content when compared to Traditional Chinese (cmn\_hant).}
	\label{fig:utf8}
\end{center}
\end{figure}

Beyond character measurements, we can also measure space more traditionally using the number of binary digits (or bits, eight of which form a byte) needed to store the same information content in different languages. The number of bits needed to store a character depends on the encoding scheme used. The most popular multilingual encoding scheme in use today is UTF-8, which uses a variable number of bytes to store a character. Almost all the characters used in English along with common punctuation marks require 8-bits (one byte) to store, while most characters used in Japanese and Chinese require 24-bits (three bytes) to store. Results from both our corpora (Figure \ref{fig:utf8}) show that information in all our languages can be stored in similar amounts of space using UTF-8 encoding. While text in English requires more characters, the number of bits in UTF-8 required to store those characters is similar to the number of bits needed to store the same content in Chinese or Japanese.

We next turn to the questions of how much is typed and how much is said in microblog posts of different languages by different types of organizations. In answering these questions, we use the character calculations from the TED talk corpus. The number of characters used in a more universal measure that is directly apparent to the user with a larger effect upon the user experience.\footnote{In contrast, the number of bits needed to store text depends on the character encoding used. Even within UTF-8, there are multiple ways to store accented characters and other characters such that two pieces of text that appear identical to the user could actually require a different number of bytes to be stored in UTF-8.}
Character limits are also the type of limits most users of microblogging platforms are familiar with since Twitter's  limit is set this way. We use the measures calculated from the TED talk corpus rather than the UDHR corpus as a more conservative estimate of the differences between languages. Furthermore, the type of language used in microblog posts is likely closer to the language used in the TED talks than to the language used in the UDHR.
Using Chinese with simplified characters (cmn\_hans) as the baseline, we have the following ratios:

\begin{equation}\label{eq:engratio}
ratio(eng,cmn\_hans)=3.21
\end{equation}
\begin{equation}\label{eq:jpnratio}
ratio(jpn,cmn\_hans)=1.30
\end{equation}

\subsection{How much is typed and posted: Organizations in action}
We find that different organizations on Weibo and Twitter platforms post messages of slightly different character lengths in different languages.

\begin{figure}
\begin{center}
	\includegraphics[width=\columnwidth]{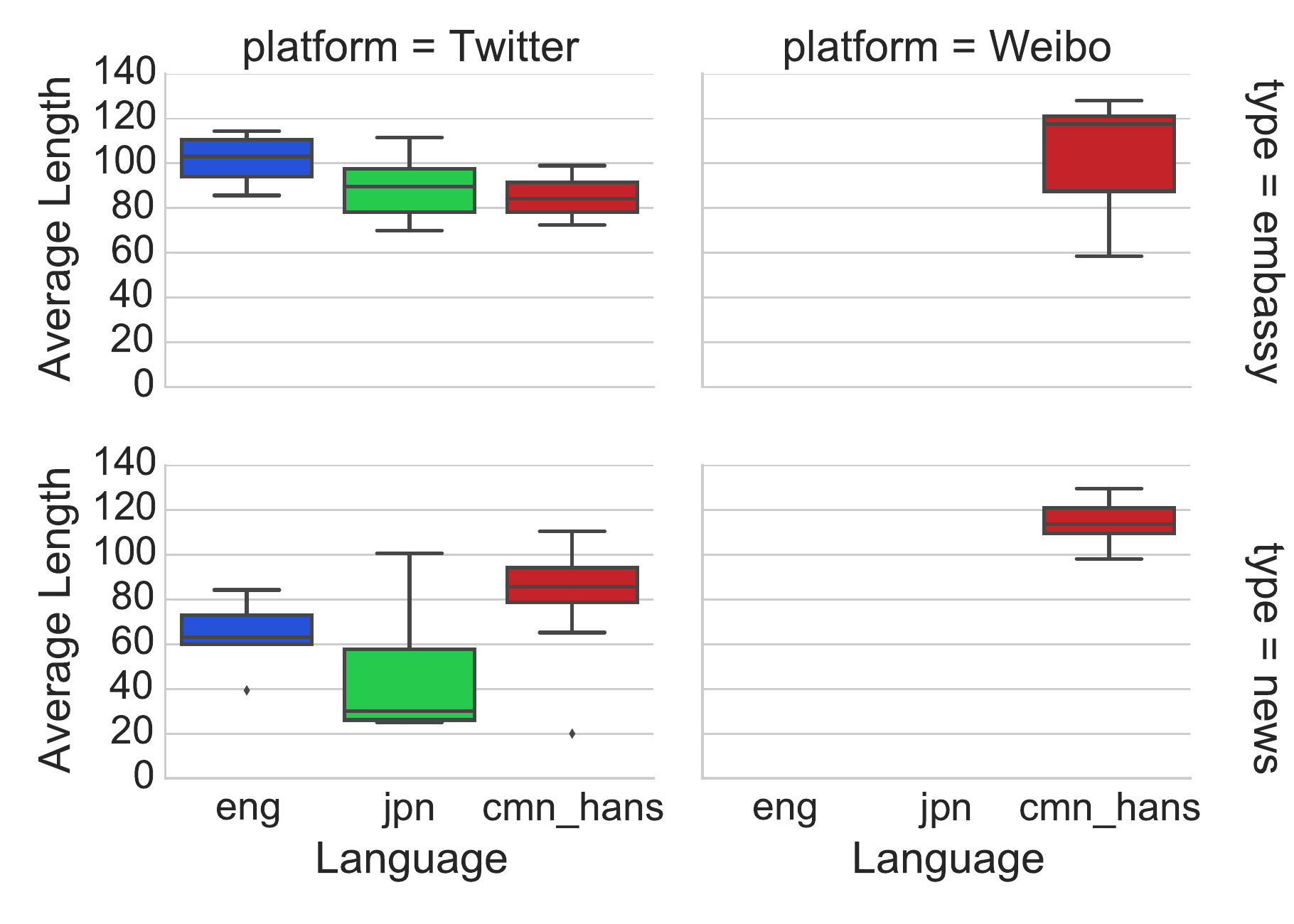}
	\caption{Length of microblog posts in characters (excluding URLs) in English (eng), Japanese (jpn) and Simplified Chinese (cmn\_hans).}
	\label{fig:microblog_chars}
\end{center}
\end{figure}

Figure \ref{fig:microblog_chars} summarizes the overall outcomes of the microblog post length (in characters) with box plots. The left subplots show the results for Twitter, and the right subplots show the results for Weibo. The top subplots show the results for embassies, and the bottom subplots show the results for news organizations.

The average lengths of posts by English-language Twitter accounts (105 characters with URLs; 81 without URLs) are close to the ``ideal length'' of a tweet reported by marketing\slash{}engagement companies of 100 or 71--100 characters \cite{lee_proven_2014,track_social_track_2012}. For English, the embassy-type accounts on average have longer posts than the news-type accounts.

The average lengths of Japanese-language Twitter accounts show a large variance, particularly within embassy-type organizations. Although this observation should not be taken as conclusive because of the limited number of Japanese-language embassy accounts ($N=18$), it is nonetheless indicative to see the relatively wide variation among the embassy-type accounts. In addition, within the news-type accounts, some Japanese-language tweets are shorter than both English and Chinese.

The average lengths of posts by Chinese-language Twitter accounts tend to be shorter than English-language posts for embassy-type organizations, whereas they tend to be longer than both English and Japanese posts for news-type organizations. The shortest average length of all the accounts in our dataset belongs to a Chinese-language Twitter account used by Falun Gong (screen name ``djy\_cn'') to broadcast news in mainland China. Comparing Twitter and Weibo in Chinese, we find that the average lengths of posts by Chinese-language Weibo accounts tend to be longer than their Twitter counterparts.

\subsection{How much is said: Comparing languages and organizations}
\begin{figure}
\begin{center}
	\includegraphics[width=\columnwidth]{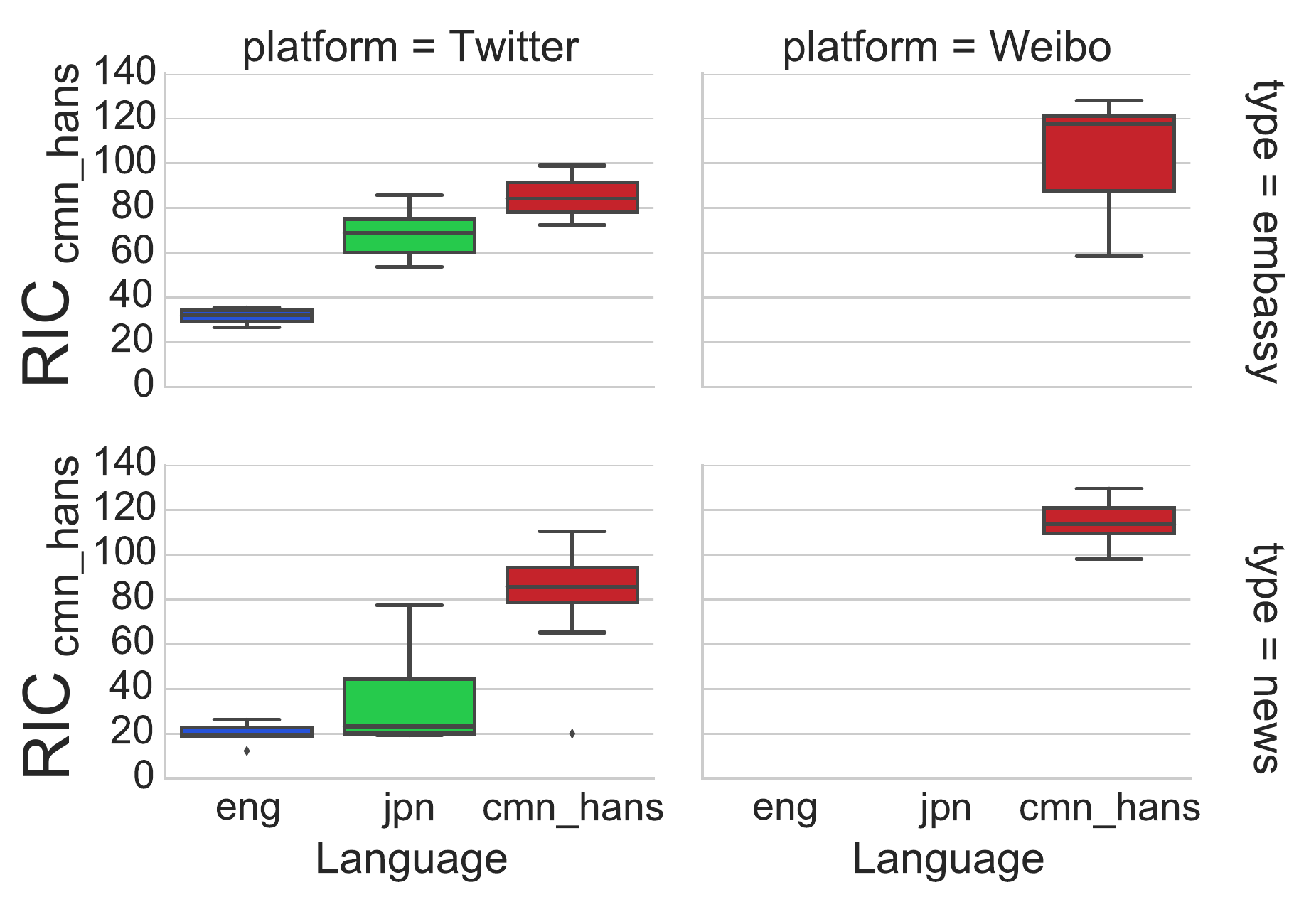}
	\caption{Relative information content (RIC) of microblog posts in English (eng), Japanese (jpn) and Simplified Chinese (cmn\_hans). RIC is shown here as the equivalent number of Simplified Chinese characters ($RIC_{cmn\_hans}$).} 
	\label{fig:microblogs_info_content}
\end{center}
\end{figure}

Combining the two sets of findings above, this final subsection derives the amount of information content in posts in different languages by different types of organizations. Similar to Figure \ref{fig:microblog_chars} in layout, Figure \ref{fig:microblogs_info_content} shows the estimated information content using the Chinese language (with simplified characters) as the baseline.

Effectively, what distinguishes Figure \ref{fig:microblogs_info_content} from Figure \ref{fig:microblog_chars} is that the number of characters in English and Japanese posts have been divided by the information content ratios found using the TED talk corpus (Equation \ref{eq:engratio} and Equation \ref{eq:jpnratio} respectively).

In contrast to the average length in characters, the comparisons of information content show a clear pattern. Most Chinese language microblog posts contain more information than either posts in Japanese or English. Furthermore, posts on Sina Weibo often contain even more content than posts on Twitter.

We see a larger variation among new organizations in comparison to embassies. Reflecting the variance in character lengths, the variance in information content per post among Japanese news organizations is quite large.
By manually inspecting the data, we found that the shorter posts often included only a news headline while the longer posts often contained a short summary of the news story.

Across the three languages, the findings suggest a general pattern where Chinese-language posts contain more information per post than either Japanese-language or English-language posts.
This difference is strongest for embassies (the top two subplots in Figure \ref{fig:microblogs_info_content}).
For the news organizations, a higher degree of overlap exists. For example, there are Chinese-language Twitter accounts such as Falun Gong (shown as an outlier dot in the figure) posting short messages.
The variation in Japanese-language accounts means that some Japanese-language accounts post messages with more information content than some Chinese-language accounts, but other Japanese-language accounts post messages with an amount of information content that is similar to many English-language accounts.

Chinese-language posts are generally longer and contain more information on Weibo than Chinese-language posts on Twitter. Confusingly, we found that some Weibo posts contained more than 140 characters whereas no Twitter posts did. As explained previously, based on further manual examination it appears that Sina Weibo enforces a byte-length limit using a variable-length character encoding  such as GBK. We found that on Sina Weibo that we were able to post messages using up to 140 Chinese characters or 280 English characters. This difference in characters available to users is an important distinction and one that our data show many Weibo users make use of as the inclusion of URLs and other English/Latin characters leaves more space for more text (of any language) on Weibo in comparison to Twitter.

\section{Discussions and Conclusions}

Based on a corpus linguistic approach with open, crowd-sourced translation data, we have updated the cross-lingual ratios of information content between English, Chinese, and Japanese. Then we have applied the results to measure empirically how much can be said and how much is actually said in microblog posts of different languages on Twitter and Sina Weibo.

The construction of relative information content measures allows researchers to move beyond length-based comparisons for microblogging to consider other content. Wikipedia, for instance, maintains a list of the one thousand most important articles and compares their lengths across language editions using ``language weights'' (relative to English) \cite{meta2015}. Our findings analyzing both the UDHR and TED talks show the impact of corpus selection for determining such weights, and calculating accurate weights will require a parallel corpus with a high degree of similarity to the type of text commonly found in Wikipedia.

We find differences in microblogging activity most strongly by platform and language, but also by organization type. In general, English-language posts use more characters than either Japanese- or Chinese-language posts on Twitter. However, once the information content per character of each language is taken into account, the relative information content per post shows that English-language posts actually contain less information per post than either Japanese- or Chinese-language posts.
As a consequence, the very definition of what constitutes ``micro'' on each platform differs by language, and this suggests that the user experience of microblogging platforms may differ greatly between languages. We further find that information content differs between Twitter and Weibo, with posts on Weibo generally containing more information than posts on Twitter. This platform difference is likely a consequence of how the platforms enforce their length limits. Twitter enforces a limit of 140 characters without regard to the storage requirements of the characters, while Weibo enforces a byte-limit.

Our work thus adds to the existing scholarship [e.g., \cite*{hong2011,mina2014,pfeil2006}] showing how the user experience of the same platform differs for users writing content in different languages. While character or byte limits impose a superficially similar measure of length across languages, we find considerable difference in the actual amount of information content available and commonly included within microblog posts written in different languages.
Hence such a superficial measure of length actually results in wide user-experience differences. These user-experience differences impact both content consumption and content creation as demonstrated by the literature on the impact of length restrictions on SMS messages [e.g., \cite*{grinter2001,ling_text_2007}].
Our findings have important ramifications for efforts to translate content across languages: content that fits within one Japanese- or Chinese-language post may not fit within one English-language post when character limits are imposed. Our findings also suggest that if the ``ideal'' length to maximize engagement sought after by marketing companies does exist, it is almost certainly language dependent.

The reliance on certain language-dependent properties of information content is bound to raise a fundamental question about how such design parameters can be uncritically applied to other languages and platforms. It also challenges the perceived wisdom such as the ``ideal length'' of tweets on the grounds that these previous findings cannot be generalized straightforwardly to other languages and/or platforms. Our findings show clear differences between the information content commonly contained in Japanese-, Chinese-, and English-language posts.

From the perspective of multilingual Internet or Internet linguistics, digital support for the main East Asian languages is an important milestone for the internationalization of the Internet. As put by Nakayama Shigeru, a major East Asian Science Technology and Society scholar \cite{chan_digital_2001}:

\begin{quote}
East Asians are accustomed to dealing with a multi-byte system, in contrast to Western mono-byte reductionist culture. It may be that in the future our multi-byte culture will prove advantageous for dealing with complex systems (p.~12).
\end{quote}

Although one does not have to agree with Shigeru's criticism of ``Western mono-byte reductionist culture,'' the findings here do suggest that platform designers and researchers need to carefully analyze what settings and assumptions may be language-specific.

By using open and freely available user-generated translation data as parallel corpora and by collecting both Twitter and Weibo posts in three languages, this research has investigated the language-specific effects of character limits on microblogging. It is expected that more systematic measurement and more linguistically diverse data sets will help both researchers and designers reexamine some of the designs and practices that are in reality language-dependent and/or language-biased and thereby find ways to account for them and develop better designs and research that are language-aware and/or language-neutral. Further research is necessary in this area to expand our knowledge of internationalization in Web Science and Internet research as well as cross-cultural Human-Computer Interaction (HCI).

Our work has been the first to investigate character lengths, byte lengths, and information content across two major microblogging platforms. Future work will build upon the work presented here to increase sample sizes, language coverage, and the types of users included. The use of less formal, more natural parallel corpora (i.e., TED talks) rather than formal legalistic prose for understanding informal conversation on Twitter and Weibo is also an important contribution. As the efforts to build parallel corpora from Twitter and other user-generated content platforms [e.g., \cite*{dalton2012,mohammadi2010,smith2010}] and from the Web more generally [e.g., \cite*{li_mining_2008,resnik_web_2003}] improve and organizations such as Meedan and Global Voices continue facilitating the human translation of user-generated content, we will have additional tools and corpora through which to examine the impact of length and other constraints on Internet-mediated communication.

\section*{Acknowledgments}
This research was supported in part by the University of Oxford's ESRC Impact Acceleration Account and Higher Education Innovation Fund (HEIF) allocation. The authors would like to thank our anonymous reviewers for their helpful comments.

\bibliographystyle{abbrv}

\end{document}